# SPARSE CHANNEL SAMPLING FOR ULTRASOUND LOCALIZATION MICROSCOPY (SPARSE-ULM)

**Erwan Hardy[1], Jonathan Porée[1], Hatim Belgharbi[1], Chloé Bourquin[1], Frédéric Lesage[2,3], Jean Provost[1,3]**

[1] Engineering Physics Department, Polytechnique Montréal, Montréal, Canada
[2] Electrical Engineering Department, Polytechnique Montréal, Montréal, Canada
[3] Montréal Heart Institute, Montréal, Canada

E-mail: jean.provost@polymtl.ca

## Abstract

Ultrasound Localization Microscopy (ULM) has recently enabled the mapping of the cerebral vasculature in vivo with a resolution ten times smaller than the wavelength used, down to ten microns. However, with frame rates up to 20.000 frames per second, this method requires large amount of data to be acquired, transmitted, stored, and processed. The transfer rate is, as of today, one of the main limiting factors of this technology. Herein, we introduce a novel reconstruction framework to decrease this quantity of data to be acquired and the complexity of the required hardware by randomly subsampling the channels of a linear probe. Method performance evaluation as well as parameters optimization were conducted in silico using the SIMUS simulation software in an anatomically realistic phantom and then compared to in vivo acquisitions in a rat brain after craniotomy. Results show that reducing the number of active elements deteriorates the signal-to-noise ratio and could lead to false microbubbles detections but has limited effect on localization accuracy. In simulation, the false positive rate on microbubble detection deteriorates from 3.7% for 128 channels in receive and 7 steered angles to 11% for 16 channels and 7 angles. The average localization accuracy ranges from 10.6 µm and 9.93 µm for 16 channels/3 angles and 128 channels/13 angles respectively. These results suggest that a compromise can be found between the number of channels and the quality of the reconstructed vascular network and demonstrate feasibility of performing ULM with a reduced number of channels in receive, paving the way for low-cost devices enabling high-resolution vascular mapping.

## 1. Introduction

By locating the centroids of sparse scatterers circulating in the vascular network, ultrasound localization microscopy (ULM) allows to go beyond the limits of conventional ultrasound imaging fixed by diffraction, and to go down to a resolution of only a few microns, using microbubbles (MB) (Errico *et al* 2015, Couture *et al* 2011, Christensen-Jeffries *et al* 2015, Desailly *et al* 2013, O'Reilly and Hynynen 2013, Couture *et al* 2018, Siepmann *et al* 2011, Viessmann *et al* 2013), or sono-activated nanodroplets (Zhang *et al* 2018, 2019). In addition to its high imaging rate, low cost, non-invasiveness and non-ionization, this modality is, as of today, the only one capable of imaging the entire vasculature of an organ within a wide field of view and in depth. Recent applications of ULM include the mapping of tumour vasculature, for early stage detection (Lin *et al* 2017), characterization (Opacic *et al* 2018), or treatment monitoring (Ghosh *et al* 2017). Other fields of interest include the detection and monitoring of treatment for cardiovascular or neurodegenerative diseases. For instance, Hingot *et al*. have imaged the cerebral perfusion of mice before, during and after ischemic strokes to evaluates the outcomes and the responses to treatment (Hingot *et al* 2020).

Numerous studies have already been carried out on sparse ultrasound imaging to reduce the acquisition time, the amount of data or the hardware complexity. Compressed sensing (Candès *et al* 2006, Candès and Romberg 2007, Donoho 2006) has shown its effectiveness in magnetic resonance imaging (Lustig *et al* 2007), photo-acoustic imaging (Provost and Lesage 2009), and X-ray tomography (Chen *et al* 2008). In ultrasound imaging, reconstructing sparse radiofrequency (RF) raw data in a wavelets base (Friboulet *et al* 2010, Liebgott *et al* 2013, Liu *et al* 2017), in the Fourier Domain (Liebgott *et al* 2013), wave atom base (Friboulet *et al* 2010, Liebgott *et al* 2013, Ramkumar and Thittai 2020) or dictionary learning base (Lorintiu *et al* 2015) have been shown. Other studies considered the sparsity of post-beamformed RF images (Achim *et al* 2010, Basarab *et al* 2013, Chernyakova and Eldar 2014, Dobigeon *et al* 2012, Quinsac *et al* 2012), scatterers distribution (David *et al* 2015, Schiffner *et al* 2012, Wagner *et al* 2012, Wang *et al* 2014, Zhang *et al* 2013) or used the sparsity of the vascular structure (Bar-Zion *et al* 2018). In the context of ultrasound, the goal was to reduce either the number of pulses/echoes, especially for synthetic transmit aperture, the number of channels or the number of samples.

Other approaches based on sparse arrays have also been proposed. Korukonda *et al.* showed the feasibility of synthetic aperture elastography imaging with a sparse array (Korukonda and Doyley 2011), decreasing the number of transmits to maintain a high frame rate. Several groups have also worked on optimizing the location of matrix array elements on several criteria : contrast, resolution, location and amplitude of the side lobes (Austeng and Holm 2002, Davidsen *et al* 1994, Diarra *et al* 2013, Roux *et al* 2018, Sciallero and Trucco 2015, Yen *et al* 2000). Relatedly, Alles *et al.* have compared different element density laws of a linear probe with their equivalent in apodization to demonstrate the interest of a non-uniform pitch (Alles and Desjardins 2020).

With regard to ULM, the feasibility of a sparse sampling of ultrasound probes has been shown in vitro in 2D with a model-based reconstruction method (Vilov *et al* 2020), and in 3D, keeping only half of a 1024-element matrix probe (Harput *et al* 2018). Moreover, as demonstrated by Desailly *et al.* in 2015, the standard deviation of the precision of localization of a unique scatterer is inversely proportional to the square root of the number of active receive channels at constant signal-to-noise ratio (SNR) (Desailly *et al* 2015). These elements seem to indicate the possibility to localize with subwavelength accuracy the microbubbles with few receive channels, and to transfer the complexity of the acquisition system to the software, reducing the costs of the ultrasound scanners, as well as the quantity of data to be collected, transferred, stored and processed.

Herein, we propose a novel sparse reconstruction framework to reduce the number of acquisition channels by randomly subsampling the receive channels, from 128 down to 16. The effects of the subsampling as well as the position of the withdrawn channels and the number of steered angles were investigated in physio-realistic simulations and *in-vivo* data, acquired in a rat brain.

## Methods

### Localization Microscopy Pipeline

The pipeline we used is conventional and similar to other approaches described in the literature (Christensen-Jeffries *et al* 2020). The reconstruction of the data was achieved by a Delay and Sum algorithm (DAS) (Montaldo *et al* 2009) on an orthonormal grid with $\lambda/4$ resolution (25.7 µm) with $\lambda$ the wavelength (see Fig. 1). To artificially decrease the microbubble concentration, and consequently increase the image quality, we separated the ascending and descending microbubbles in *in-vivo* data. To do so we applied a filter on the Fourier transform of a pixel signal (Osmanski *et al* 2012). This method was previously shown as effective by Huang et al (Huang *et al* 2020). For *in-vivo* data, a Singular Value Decomposition (SVD) (Demené *et al* 2015) was computed to reject the tissue signal by withdrawing the first 27 eigenvectors. Then, the point spread function (PSF) of a microbubble located in the center of the reconstructed region was simulated considering a fully populated array and correlated with the reconstructed images. A Gaussian fitting was then performed on the correlation maps to localize microbubbles centers with a subwavelength precision. The microbubbles were tracked in time using a nearest-neighbour criterion to eliminate microbubbles that did not persist for more than two consecutive frames.

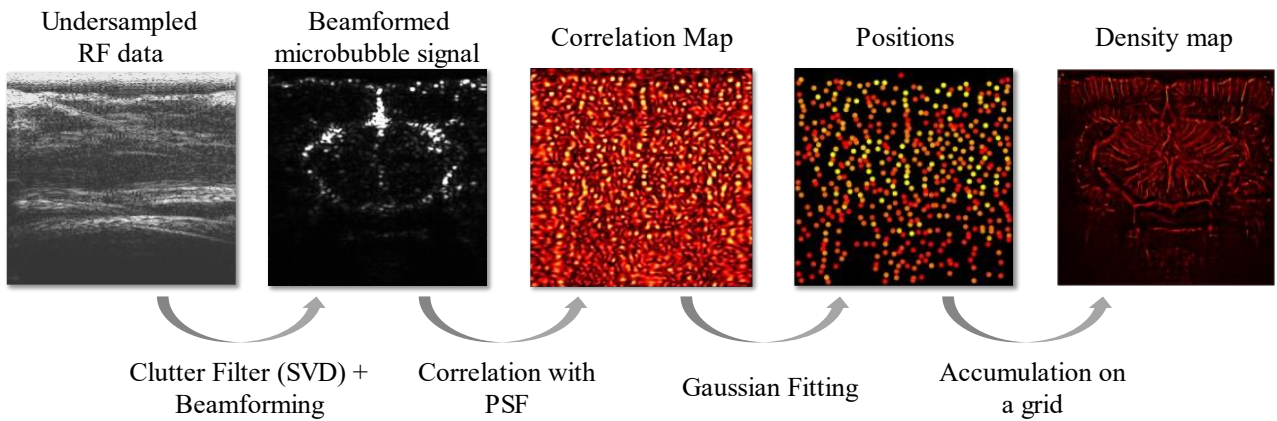

**Figure 1 – Ultrasound localization microscopy pipeline.** The RF data containing the microbubble and tissue signal was reconstructed from one spatial axis (x) and one temporal axis to two spatial axes (x,z). A SVD was performed on the raw or beamformed data to remove the tissue signal. A correlation was then performed, and a Gaussian fitting on the pixels with the highest correlations allowed us to obtain the position with sub-pixel precision. These positions were accumulated on a grid to form the angiogram.

### Receive Channel Reduction Method

The approach proposed herein to decrease the number of acquisition channels consists in

1) randomly selecting groups of receive elements, from 128 to 16 elements, for each insonification angle, which

2) remain constant during an entire buffer. The latter aspect is central to the approach, as it enables the use of standard SVD filtering to isolate microbubbles (see Fig. 2).

For instance, a buffer containing 2000 frames when using 5 insonification angles is split into 5 groups of 400 frames according to the insonification angle. In each group, a single set of active elements is randomly selected. Since each group is independent, so some elements may be present several times, and others may be absent.

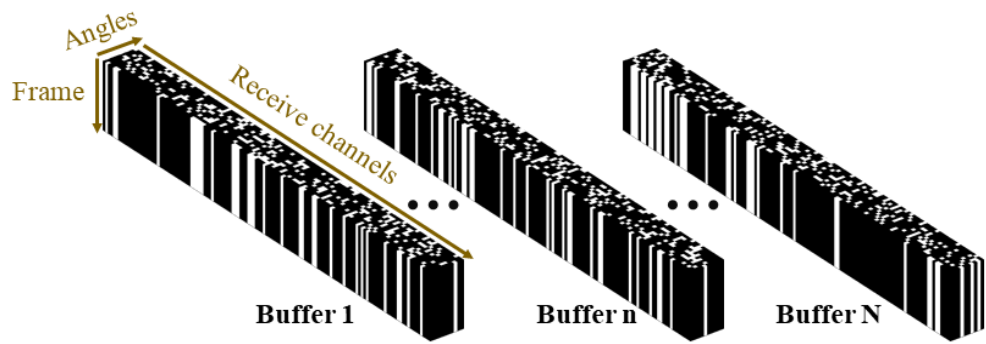

**Figure 2 - Sub-sampling method.** Diagram showing an example of the subsampling method used in the 3D space formed by the frame number (slow time), emitted angles, and receive channels. Active channels are represented in white. The selected active channels change from one angle to the other, but stay constant from one frame to the next within a buffer.

To evaluate the influence of the elements position in the reconstructions, several probability laws were tested. The elements were then selected among the 128 of the probe used with a higher probability for the external elements (Ext1 and Ext2 laws), for the central elements (Cen1 and Cen2 laws) or in a uniform way (Uni law).

### Creation of the In-Silico Phantom and Metrics Extracted

Six vascular networks from mouse brains were imaged using 2-photon imaging and reconstructed (Damseh *et al* 2019). The vasculature was then segmented to circulate microbubbles following a physio-realistic distribution. The microbubble positions thus obtained were dilated by a factor 5 to fill the field of view of the probe, and cut along one direction. The slices were translated and rotated to fill a field of view equivalent to that of a rat brain while avoiding redundancy. The positions of 13 microbubbles per slice in the resulting phantom were simulated to obtain ultrasound images with a frame rate of 1000 frames per second (fps), 13 angles (from -3° to 3° in steps of 0.5°) and a concentration of 3.84 microbubbles/mm³ which was found to be an optimal concentration for our localization algorithms and ultrasound probe. To emulate the linear acoustic response of the microbubbles, we used an in-house GPU implementation of the frequency-based simulation software (Shahriari and Garcia 2018). It has been set up to emulate a L-22-14 probe (Vermon, France) at 15 MHz. 50 buffers of 400 frames were obtained and stored as in-vivo data to be reconstructed (see Fig. 3). A Gaussian noise of 33dB was added to the RF data, in order to have a noise level of 10dB on the beamformed data with one angle.

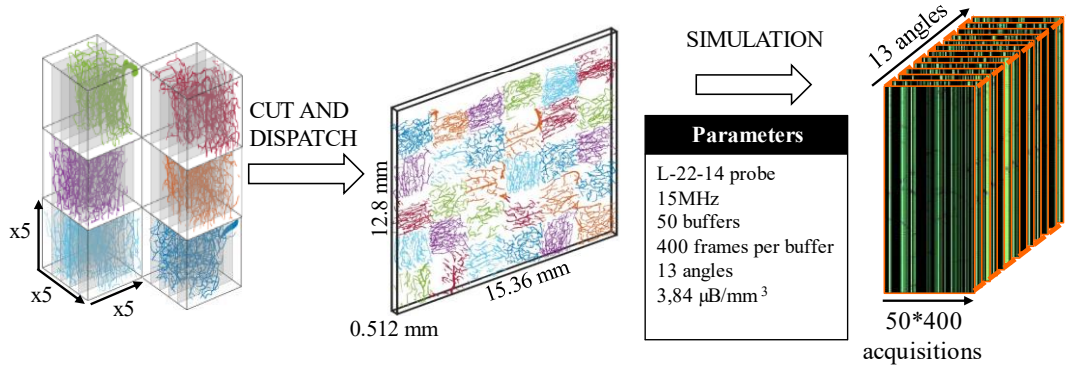

**Figure 3 – Physio-realistic phantom pipeline.** At the left, the six segmented angioarchitectures from 2-photon measurement that were used to generate microbubble physio-realistic position and speed. At the center, the obtained phantom. At the right, the simulated sparse RF data.

The microbubbles were then located and uniquely matched with the positions used for the simulation. Three metrics were measured:

- the false positive rate (FPR) defined as

$$\frac{\text{number of unmatched localized microbubbles}}{\text{number of localized microbubbles}} \tag{1}$$

- the false negative rate (FNR) defined as

$$\frac{\text{number of undetected simulated microbubbles}}{\text{number of simulated microbubbles}} \tag{2}$$

- the mean distance between the simulated and localized microbubbles, which will be referred to as accuracy

The standard deviation of the distances between the simulated and localized microbubbles will be referred to as precision. All the metrics, including the STD, were calculated per frame.

This phantom has been designed to fill the entire probe's field of view with a homogeneous distribution of microbubbles in it, in order to avoid evaluation bias due to a higher number of active receive elements above denser microbubble zones.

*In-Vivo Acquisition Setup*

The acquisition was performed on a female rat's brain after craniotomy sedated with Isofluorane (2 %) and placed on a monitoring platform (Labeo Technologies Inc., Montréal, Canada) to monitor its heart and respiratory rate. The platform was heated to maintain the body temperature at 35°C. Three steered plane (-1, 0 and 1°) were emitted with a fully populated array (L22-14, 18 MHz, Vermon, France) and backscattered signals were recorded with a Vantage 256 system (Verasonics, WA, USA) after a bolus injection in the tail vein of a 50-µL MB solution ($1.2 \times 10^{10}$ microbubbles per milliliter, Definity, Lantheus Medical Imaging, Billerica, MA, USA) diluted in 50 µL of saline. Each acquisition consisted of blocks, that will be referred as buffers, of 400 RF data, acquired at an imaging cadence of 1000 frames per seconds.

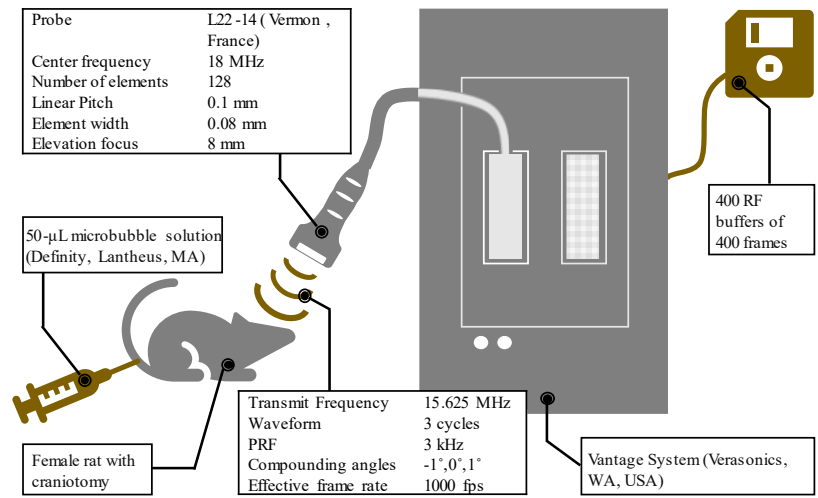

**Figure 4 – *In-vivo* acquisition setup**. Diagram presenting the parameters and equipment used, from acquisition to data recording. The frame containing the RF data for the 3 angles are arranged by buffers of 400. 400 buffers are acquired, at a rate of one every 2 seconds approximately.

### Angiogram Display

To display the angiogram, the positions of microbubbles having a correlation above a threshold were accumulated on a grid of $\lambda/12$ (8.6 µm). The threshold was set at 0.5 for *in-vivo* data and determined by the Otsu method for in *in-silico* data. The Otsu method was chosen through the realization of ROC curves on *in-silico* metrics. A median filter with a 2 x 2 kernel (8.5 µm x 8.5 µm) was performed. The microbubble density was displayed on a logarithmic scale along with a gamma correction of 2 to ease the visualization of small vessels in *in-vivo* density maps.

### Parameters of Interest

Three main parameters have been studied in this work: the number of channels in receive, the number of angles, and the position of the elements in receive. The first two have been studied together, while the last one has been studied for 32 receiving channels and 5 angles.

The impact of the amount of data was also observed. For this, the number of data buffers was proportionally adjusted downward to compensate for a higher number of channels: 400 for 16 elements, 200 for 32 elements down to 50 for 128 elements.

## Results

### In-Silico Results

### Microbubbles can be accurately localized using an under-sampled probe

Several parameters influence the quantity of data and the required transfer rate: sampling frequency, acquisition depth, number of channels, number of pulse echoes. We have chosen to vary the number of angles emitted and the number of channels in order to find out if a compromise could be found between these parameters, on the one hand, and the image quality, on the other hand. The *in-silico* results of this study are presented in Figure 5. The FPR, linked to the contrast of the image, decreased with the increase in number of channels to converge towards 3.5 % for each number of angles. However, with 16 channels, the FPR exhibited important differences: 26 %, 7.7 % and 4.2 % of false positive microbubbles in average for 3, 7 and 11 angles, respectively. The FNR is also linked to the contrast, a low FNR meaning a high level of detection of the microbubbles. Simulations with only 3 angles stood out with a FNR of 79 % for 3 angles and 16 receive channels. Nevertheless, the accuracy of localization was similar for each number of angles and channel, with a variation smaller than 2 µm. The average accuracy was 10.9 µm and 9.05 µm for 16 channels/3 angles and 128 channels/9 angles respectively. The best accuracy is not achieved for 13 angles but for 9 angles at 128 channels. One hypothesis is that with the calculated correlation threshold, more microbubbles are considered in the calculation of the metrics, with a greater distance from the simulated microbubble. This hypothesis would be in agreement with the fall of the FNR for these same values of angles and channels. It is important to note that this variation is small with respect to the precision, which is of the order of 9 µm. The mean localization accuracy was approximately 10 µm or λ/10.

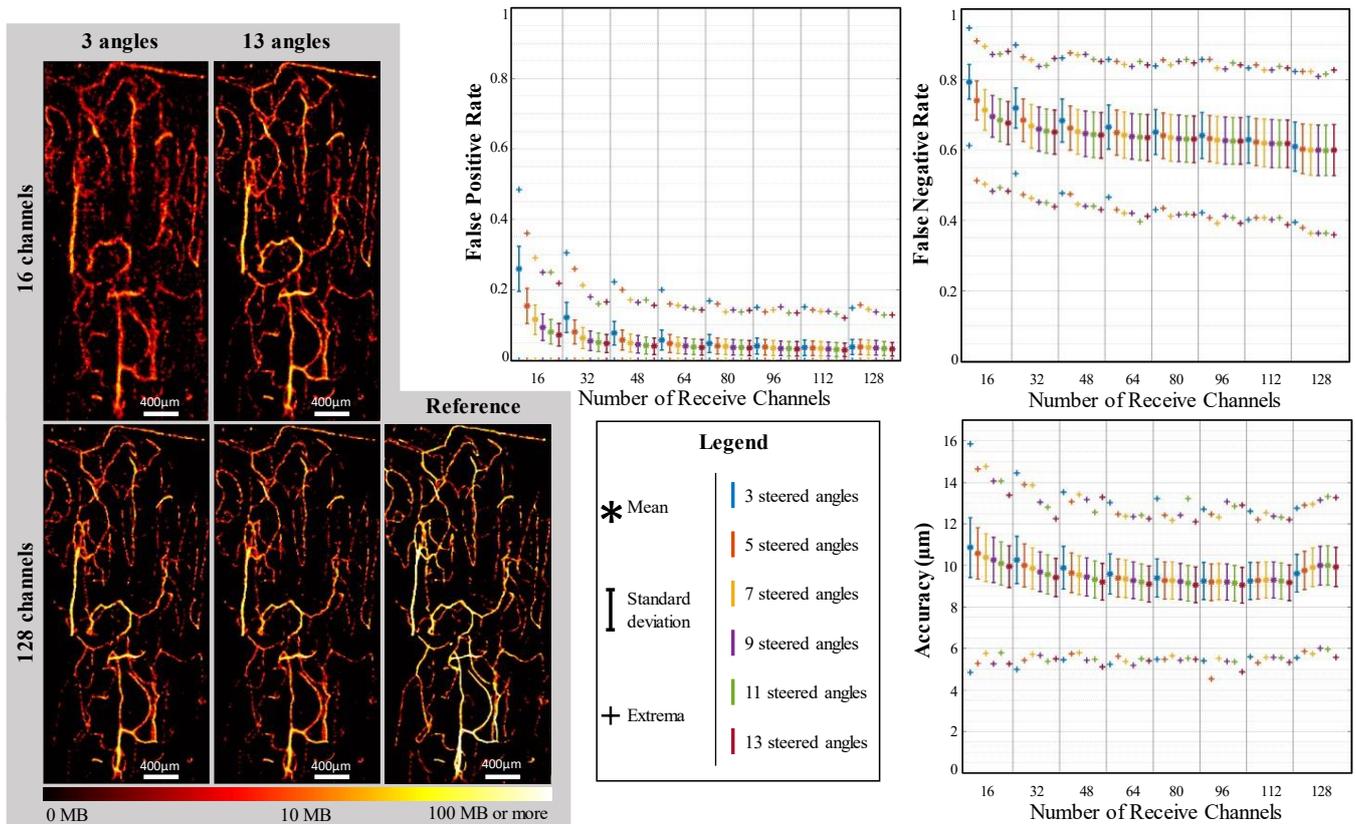

**Figure 5 – *In-silico* results for different numbers of active receive channels and angles.** FNR, FPR and accuracy for a number of active receive elements varying from 16 to 128 by steps of 16 and number compounding angles varying from 3 to 13 keeping an angular sampling of 0.5°. The curves are obtained by averaging extracted metrics on the 20000 reconstructed frames. Angiograms represent a region of interest extracted from the in-silico phantom. The reference was obtained by accumulating the positions of the microbubbles while considering their azimuthal position (y) null.

### The position of the active receive elements is of little importance in silico

Since the position of the sidelobes generated by the microbubbles depends on the sampling of the probe, we have developed an approach where the active elements in receive are changed as often as possible, in order to avoid redundancy in the position of the sidelobes that can lead to false detections and artifacts on the final angiogram. Hence, instead of optimizing a deterministic configuration of active elements, we select them using different probability distributions that, e.g., promote either the central elements or the side elements. The comparison was made with 32 channels in receive and 5 compounded angles (see Fig. 6). We can see that the central elements increase the accuracy (9.4 µm in average for Cen2 and 10.3 µm for Ext2) and the precision, while the side elements enhanced the detection of microbubbles with a low FNR (71,3 % of false negative microbubbles for Cen2 and 68,4 % for Ext1). The FPR is the lowest for a uniform selection of channels (8.0 % of false positive microbubbles). However, the variations of these metrics are small compared to the STD. The angiograms show minor qualitative differences except a slightly higher intensity of the central vessels for Cen1 law and a higher intensity of the lateral vessels for Ext 1 law, which is expected.

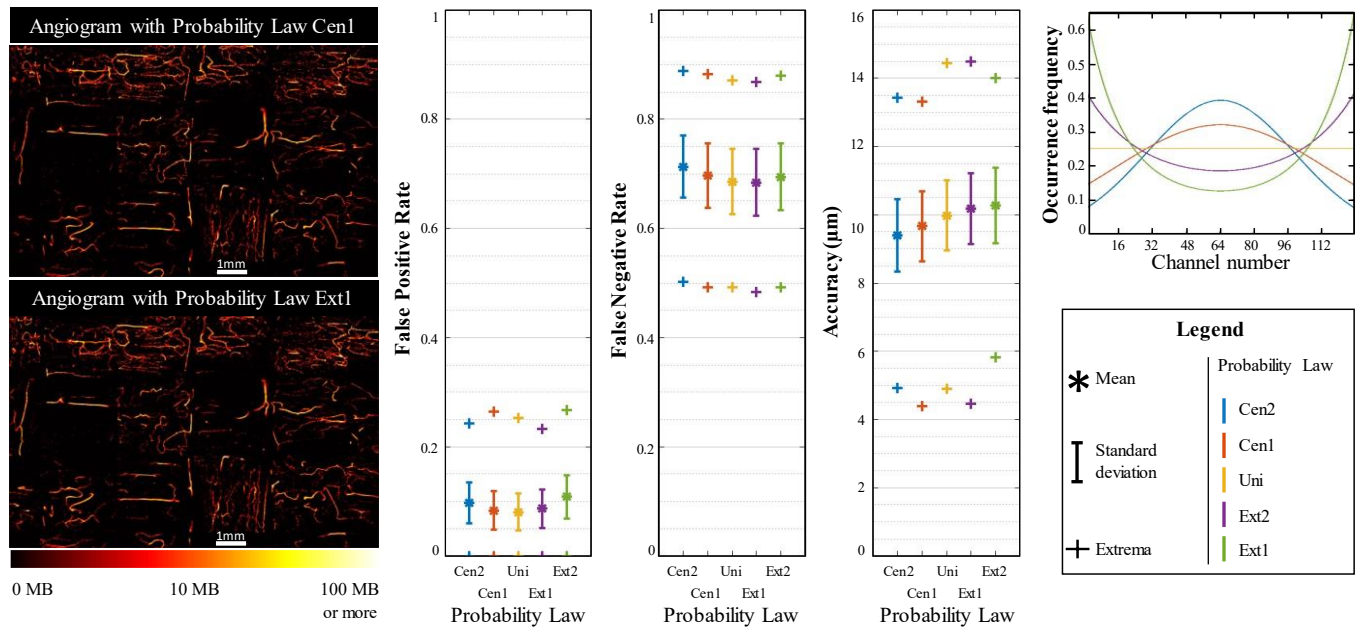

**Figure 6 –** *In-silico* **comparison of different channel selection probability laws.** Comparison of 5 probability laws for 32 receive channels and 5 angles (-1, -0.5, 0, 0.5, 1). The curves are obtained by averaging metrics on the 20000 reconstructed frames. The occurrence frequencies correspond to the theoretical ones.

### In-Vivo Results

#### False microbubble detections decrease the contrast and smaller vessels disappear

To evaluate the method in vivo, we worked on data from a rat acquisition with craniotomy. The signals of the 128 channels were recorded and subsampled during processing. In figure 7, results show that the background noise increases, and the smaller vessels tend to disappear with the decrease in number of channels. For example, the two vessels indicated by a white arrow in the center of the green region of interest, were only visible in the angiograms reconstructed using 32 channels or more. However, their distinction and even more the measurement of their width remain difficult on profiles without the fully populated array. Similarly, for the magenta region of interest, only the two main vessels are visible with 16 channels. Other vessels and their ramifications appear progressively with the addition of channels.

Signal degradation due to the reduction in the number of channels is inevitable. However, with 16 channels we obtain vessels with both better resolution and contrast than with Contrast Enhanced Ultrafast Power Doppler. Indeed, although the presence of microbubbles increases contrast, diffraction spreads the large vessels and degrades resolution. The Full Width at Half Maximum (FWHM) of the vessel on the left side of the profile thus increases from about 150 µm to more than 200 µm.

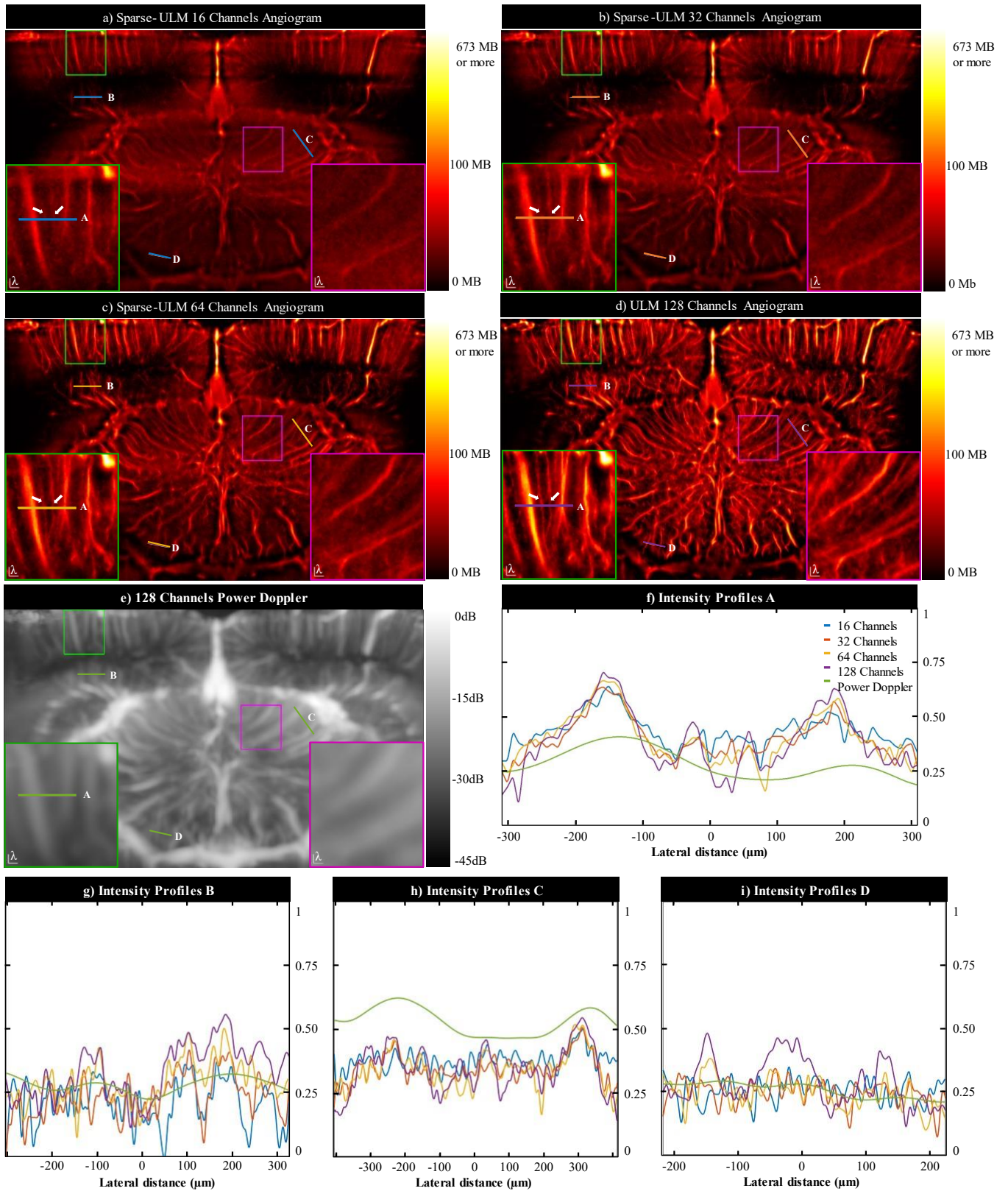

**Figure 7 – *In-vivo* results for different numbers of active receive channels**. a), b), c) and d) are respectively the angiograms obtained with 16, 32, 64 and 128 channels in receive. e) is the power Doppler obtained with the same data set, i.e. with microbubbles, displayed at -45dB. Power Doppler and angiograms are in logarithmic scale. 2 regions of interest and one profile are extracted. The profiles f), g), h) and i) are normalized between 0 and 1 with respect to their original image.

We have also studied the effect of active receive element count for a fixed amount of data transfer by increasing the number of imaging frames when using fewer receive elements (Fig. 8). Overall, the contrast increased with the number of active receive elements. However, the smaller vessels disappear, even with 128 channels. For instance, the small vessels indicated by white arrows in the green region of interest, are more easily discernible with 64 channels than with 128. In addition, the network filing is impacted with 50 buffers, as shown by the discontinuity of these vessels. Contrast and resolution of angiograms remain higher than with Power Doppler.

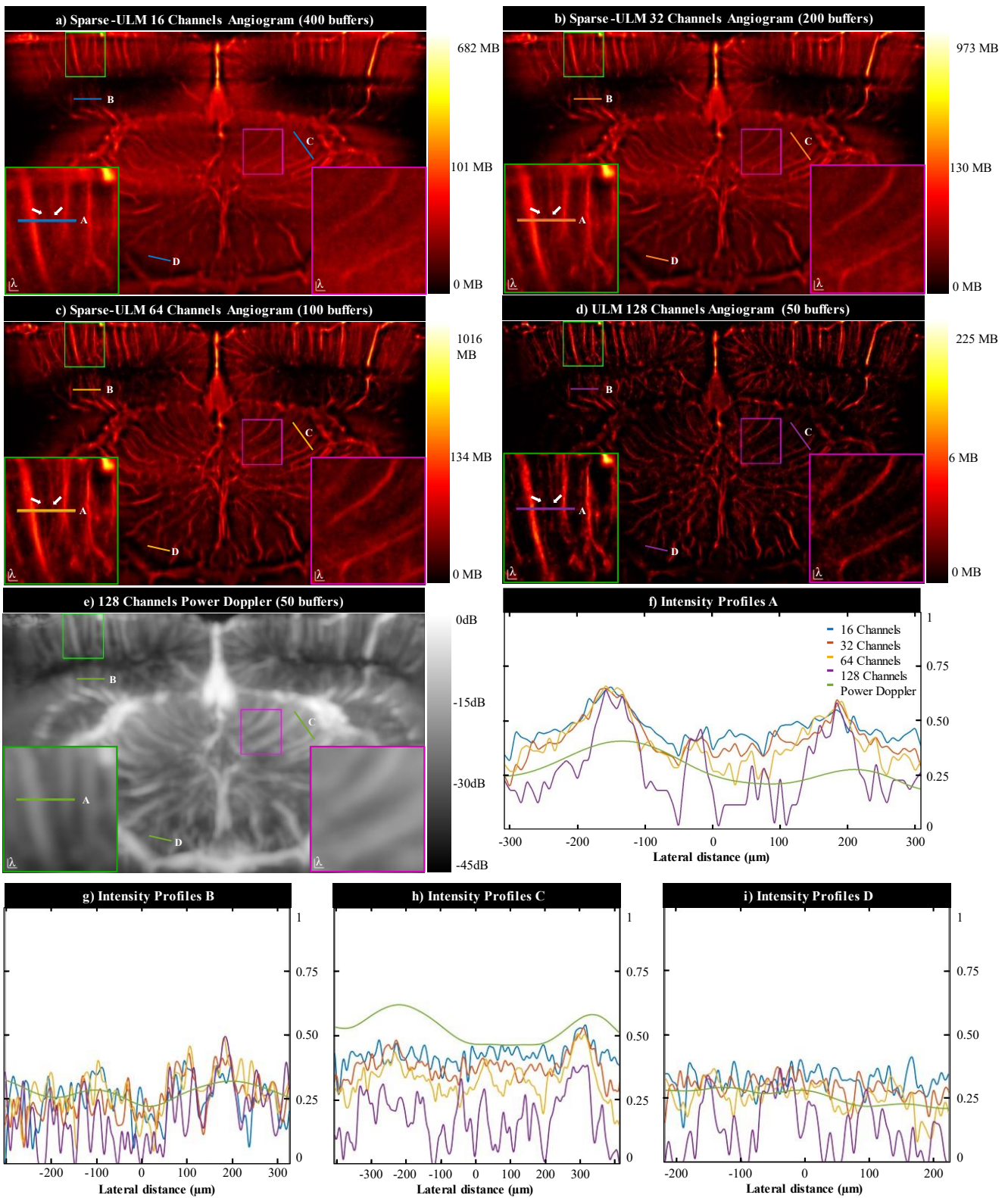

**Figure 8 – *In-vivo* results for different numbers of active receive channels with the same amount of data.** a), b), c) and d) are respectively the angiograms obtained with 16, 32, 64 and 128 channels in receive. e) is the power Doppler obtained with the same data set, i.e. with microbubbles, displayed at -45dB. Power Doppler and angiograms are in logarithmic scale and the number of buffers used to reconstruct the images are adjusted to keep the same quantity of data. Two regions of interest and one profile are extracted. The profiles f), g), h) and i) are normalized between 0 and 1 with respect to their original image.

*Slightly favoring the central elements provides better overall image contrast*

We reconstructed the in vivo data with the same laws of probability of occurrence of the elements as shown in Figure 6. The results, Figure 9, show that the degradation of the vessels in the center of the image is faster with a probability law favoring the external elements (Ext1 and Ext2) than that of the external vessels with a law favoring the central elements (Cen1 and Cen2).

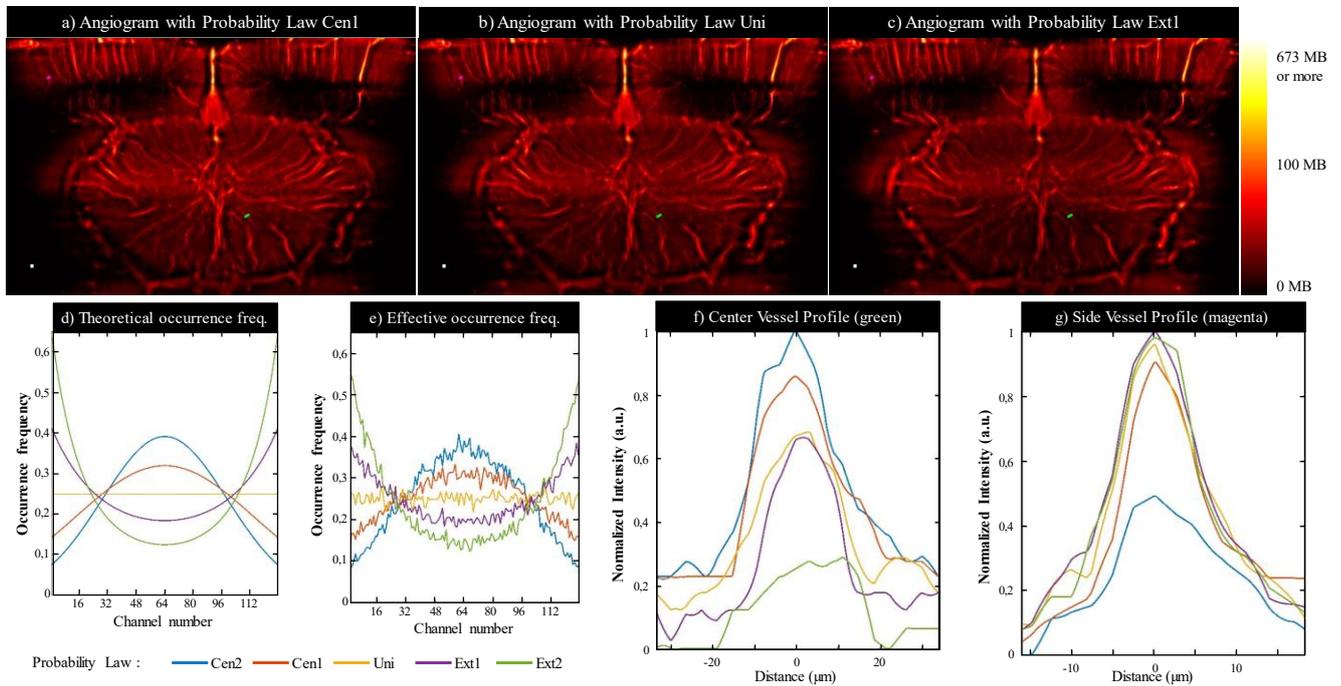

**Figure 9 – *In-vivo* comparison of different channel selection probability laws.** a), b) and c) are density maps of the in vivo rat angioarchitecture with the probability law Cen1, Uni and Ext1. d) and e) are the theoretical law for the random selection, and the effective occurrence frequency of each element. f) and g) are vessels profiles extracted from the angiogram, one at the center of the image, one outside. They are normalized between 0 and 1 with respect to the minimum and maximum of the profiles.

## Discussion

This study introduced the Sparse-ULM method to decrease the transfer rate in ultrasound scanner while achieving ULM, to decrease the hardware complexity and the cost of these devices. This method allowed us to achieve angiographic images despite a reduction of active receive elements by a factor of 8.

We have shown with the help of a physio-realistic phantom that as the number of channels in receive decreased, the number of incorrectly detected or non-localized microbubbles increased. Incorrectly localized microbubbles, or false positives, created a background noise when the undetected microbubbles, or false negatives, decreased the vessel signal, which resulted in a decrease in contrast. However, accuracy and precision were maintained with the decrease in the number of channels which means that our ability to accurately locate the microbubble center remained intact even with a degraded PSF. One of the reasons could be that, even if the shape of the PSF is modified, the main changes took place on the side lobes. As a result, the correlation with a small kernel (11pixels by 11 pixels) is only slightly impacted and detection remained good, as well as localization, where the Gaussian fitting was robust to these changes. The preservation of the precision could seem to contradict the theory of Desailly et. al, which predicts a decrease in precision proportional to the square root of the number of elements in the probe (Desailly *et al* 2015). However, our models differ on several points: here the localization is done on several microbubbles within the same image. Moreover, the decay described above is obtained by deriving the Cramer-Rao lower bound, which may not be reached. In addition, our phantom had no tissue. This limitation is important because the lower bound of Desailly et al. also depends on the SNR, and it is known that in the presence of noise, the reduction of the number of channels will have more impact on the signal quality. Other studies are therefore to be carried out, especially since the impact of the subsampling the SVD is always to be evaluated.

The study of how to choose the elements of the probe indicated that a uniform law or one that slightly promotes the central elements is to be recommended since they seemed to be a good compromise between FPR, FNR and accuracy. Nevertheless, given the small variation in metrics with respect to their STD, no conclusion was possible. The rapid convergence of the FNR and the FPR with the increase in the number of transmitting channels, especially for 5 angles and more, as well as the invariance of the accuracy and precision allow us to conclude that a compromise between the quality of the angiogram and the number of receiving channels is theoretically possible.

On in-vivo data, the decrease of the sensibility and of the amount of data impacted the filling of the angiograms, as it has been shown in figure 8. The smaller vessels are the first to disappear in both cases. The disappearance of these vessels is probably due to a lower number of microbubbles within them. According to Hingot et al., beyond the lower blood flow in small vessels, these contain a higher quantity of red blood cells in relation to the number of microbubbles, so much so that a $100 \mu m$ vessel transits 100 times more microbubbles than a $5 \mu m$ vessel. (Hingot *et al* 2019). Here we lacked data with a higher number of angles to study this parameter correctly. The main arterial structures were preserved, and more visible than with Power Doppler.

As for the in-silico part, the location of the active receive elements did not have a preponderant impact, even if the uniform laws or a law slightly favoring the center seemed to give a better angiogram overall, with more little structures in the centre. As a result, the distribution of active receive elements was not the main driver of improvement. Nonetheless, it might be interesting to investigate further on the choice of distributions between the compounded angles. Indeed, although it is necessary to change the elements as often as possible, the selected configuration has a direct impact on the location of the side lobes, and a judicious distribution of the configurations according to the angles of a frame could make it possible to limit the side lobes during the compounding, and thus reduce the false positive rate.

As far as bandwidth is concerned, a reduction in the number of active receive elements from 128 to 16 reduce the number of channels required by a factor of 8. An integration on a compact system is then fully feasible. Such a system would leave the possibility of acquiring a larger number of channels by splitting them into multiples of 16, thus making it possible to adapt the image quality to the desired application.

Overall, the results are encouraging, but further research, particularly on the possibility of extracting biomarkers from under-sampled data, will allow further conclusions to be drawn on the effectiveness of this method.

In order to further improve the method, other techniques could be integrated, based either on compressed sensing or on neural networks. The union of the two also gives interesting results. Indeed, the injection of knowledge via deep learning has allowed the reduction of artifacts caused by aliasing in MRI (Yang *et al* 2018, Lee *et al* 2017) and CT (Han *et al* 2016). To get the full potential of this method, it must be applied to 3D ULM, where the need for channel reduction is greater. The translation work is in progress, and we believe that the results presented here can be generalized to 3D imaging.

These are avenues that will not remain unexplored in ultrasound imaging.

## Conclusion

ULM makes it possible to reflect the sparsity of the medium to be imaged, directly on the hardware, by under sampling a conventional imaging probe and maintaining the ability to precisely localize the microbubbles circulating in the vessels. Although a loss of signal, due to false detections, degrades the smallest vessels, especially for a low number of emitted angles, the images obtained remain superior to what can be obtained with a traditional power Doppler. This paves the way for less expensive and more compact ultrasound scanners.

## Acknowledgement

We acknowledge the support of IVADO, TransMedTech, New Frontiers in Research Fund, and the Canadian First Research Excellent Fund (Apogée/CFREF).

## Ethics

The use and care of laboratory animals has been done in accordance with the recommendations of the Canadian Council on Animal Care and all animal studies have been approved by the Montreal Heart Institute's Animal Care Ethics Committee. (Permit Number: 2019-2464, 2018-32-03).

## References

Achim A, Buxton B, Tzagkarakis G and Tsakalides P 2010 Compressive sensing for ultrasound RF echoes using a-Stable Distributions *2010 Annual International Conference of the IEEE Engineering in Medicine and Biology* 2010 Annual International Conference of the IEEE Engineering in Medicine and Biology pp 4304–7

Alles E J and Desjardins A E 2020 Source Density Apodization: Image Artifact Suppression Through Source Pitch Nonuniformity *IEEE Trans. Ultrason., Ferroelect., Freq. Contr.* **67** 497–504

Austeng A and Holm S 2002 Sparse 2-D arrays for 3-D phased array imaging - design methods *IEEE Transactions on Ultrasonics, Ferroelectrics, and Frequency Control* **49** 1073–86

Bar-Zion A, Solomon O, Tremblay-Darveau C, Adam D and Eldar Y C 2018 SUSHI: Sparsity-Based Ultrasound Super-Resolution Hemodynamic Imaging *IEEE Trans. Ultrason., Ferroelect., Freq. Contr.* **65** 2365–80

Basarab A, Liebgott H, Bernard O, Friboulet D and Kouamé D 2013 Medical ultrasound image reconstruction using distributed compressive sampling *2013 IEEE 10th International Symposium on Biomedical Imaging* 2013 IEEE 10th International Symposium on Biomedical Imaging pp 628–31

Candès E J, Romberg J K and Tao T 2006 Stable signal recovery from incomplete and inaccurate measurements *Communications on Pure and Applied Mathematics* **59** 1207–23

Candès E and Romberg J 2007 Sparsity and incoherence in compressive sampling *Inverse Problems* **23** 969–85

Chen G-H, Tang J and Leng S 2008 Prior image constrained compressed sensing (PICCS): A method to accurately reconstruct dynamic CT images from highly undersampled projection data sets *Med Phys* **35** 660–3

Chernyakova T and Eldar Y C 2014 Fourier-domain beamforming: the path to compressed ultrasound imaging *IEEE Transactions on Ultrasonics, Ferroelectrics, and Frequency Control* **61** 1252–67

Christensen-Jeffries K, Browning R J, Tang M, Dunsby C and Eckersley R J 2015 In Vivo Acoustic Super-Resolution and Super-Resolved Velocity Mapping Using Microbubbles *IEEE Transactions on Medical Imaging* **34** 433–40

Christensen-Jeffries K, Couture O, Dayton P A, Eldar Y C, Hynynen K, Kiessling F, O'Reilly M, Pinton G F, Schmitz G, Tang M-X, Tanter M and van Sloun R J G 2020 Super-resolution Ultrasound Imaging *Ultrasound in Medicine & Biology* **46** 865–91

Couture O, Besson B, Montaldo G, Fink M and Tanter M 2011 Microbubble ultrasound super-localization imaging (MUSLI) *2011 IEEE International Ultrasonics Symposium* 2011 IEEE International Ultrasonics Symposium pp 1285–7

Couture O, Hingot V, Heiles B, Muleki-Seya P and Tanter M 2018 Ultrasound Localization Microscopy and Super-Resolution: A State of the Art *IEEE Transactions on Ultrasonics, Ferroelectrics, and Frequency Control* **65** 1304–20

Damseh R, Pouliot P, Gagnon L, Sakadzic S, Boas D, Cheriet F and Lesage F 2019 Automatic Graph-Based Modeling of Brain Microvessels Captured With Two-Photon Microscopy *IEEE Journal of Biomedical and Health Informatics* **23** 2551–62

David G, Robert J, Zhang B and Laine A F 2015 Time domain compressive beam forming of ultrasound signals *The Journal of the Acoustical Society of America* **137** 2773–84

Davidsen R E, Jensen J A and Smith S W 1994 Two-Dimensional Random Arrays for Real Time Volumetric Imaging *Ultrasonic Imaging* **16** 143–63

Demené C, Deffieux T, Pernot M, Osmanski B-F, Biran V, Gennisson J-L, Sieu L-A, Bergel A, Franqui S, Correas J-M, Cohen I, Baud O and Tanter M 2015 Spatiotemporal Clutter Filtering of Ultrafast Ultrasound Data Highly Increases Doppler and fUltrasound Sensitivity *IEEE Transactions on Medical Imaging* **34** 2271–81

Desailly Y, Couture O, Fink M and Tanter M 2013 Sono-activated ultrasound localization microscopy *Appl. Phys. Lett.* **103** 174107

Desailly Y, Pierre J, Couture O and Tanter M 2015 Resolution limits of ultrafast ultrasound localization microscopy *Phys. Med. Biol.* **60** 8723–40

Diarra B, Robini M, Tortoli P, Cachard C and Liebgott H 2013 Design of Optimal 2-D Nongrid Sparse Arrays for Medical Ultrasound *IEEE Transactions on Biomedical Engineering* **60** 3093–102

Dobigeon N, Basarab A, Kouamé D and Tourneret J-Y 2012 Regularized Bayesian compressed sensing in ultrasound imaging *2012 Proceedings of the 20th European Signal Processing Conference (EUSIPCO)* 2012 Proceedings of the 20th European Signal Processing Conference (EUSIPCO) pp 2600–4

Donoho D L 2006 Compressed sensing *IEEE Trans. Inform. Theory* **52** 1289–306

Errico C, Pierre J, Pezet S, Desailly Y, Lenkei Z, Couture O and Tanter M 2015 Ultrafast ultrasound localization microscopy for deep super-resolution vascular imaging *Nature* **527** 499–502

Friboulet D, Liebgott H and Prost R 2010 Compressive sensing for raw RF signals reconstruction in ultrasound *2010 IEEE International Ultrasonics Symposium* 2010 IEEE International Ultrasonics Symposium pp 367–70

Ghosh D, Xiong F, Sirsi S R, Mattrey R, Brekken R, Kim J-W and Hoyt K 2017 Monitoring early tumor response to vascular targeted therapy using super-resolution ultrasound imaging *2017 IEEE International Ultrasonics Symposium (IUS)* 2017 IEEE International Ultrasonics Symposium (IUS) pp 1–4

Han Y S, Yoo J and Ye J C 2016 Deep Residual Learning for Compressed Sensing CT Reconstruction via Persistent Homology Analysis *arXiv:1611.06391 [cs]* Online: http://arxiv.org/abs/1611.06391

Harput S, Christensen-Jeffries K, Brown J, Zhu J, Zhang G, Leow C H, Toulemonde M, Ramalli A, Boni E, Tortoli P, Eckersley R J, Dunsby C and Tang M-X 2018 3-D Super-Resolution Ultrasound Imaging Using a 2-D Sparse Array with High Volumetric Imaging Rate *2018 IEEE International Ultrasonics Symposium (IUS)* 2018 IEEE International Ultrasonics Symposium (IUS) pp 1–9

Hingot V, Brodin C, Lebrun F, Heiles B, Chagnot A, Yetim M, Gauberti M, Orset C, Tanter M, Couture O, Deffieux T and Vivien D 2020 Early Ultrafast Ultrasound Imaging of Cerebral Perfusion correlates with Ischemic Stroke outcomes and responses to treatment in Mice *Theranostics* **10** 7480–91

Hingot V, Errico C, Heiles B, Rahal L, Tanter M and Couture O 2019 Microvascular flow dictates the compromise between spatial resolution and acquisition time in Ultrasound Localization Microscopy *Scientific Reports* **9** 2456

Huang C, Lowerison M R, Trzasko J D, Manduca A, Bresler Y, Tang S, Gong P, Lok U-W, Song P and Chen S 2020 Short Acquisition Time Super-Resolution Ultrasound Microvessel Imaging via Microbubble Separation *Scientific Reports* **10** 6007

Korukonda S and Doyley M M 2011 Estimating Axial and Lateral Strain Using a Synthetic Aperture Elastographic Imaging System *Ultrasound in Medicine & Biology* **37** 1893–908

Lee D, Yoo J and Ye J C 2017 Deep residual learning for compressed sensing MRI *2017 IEEE 14th International Symposium on Biomedical Imaging (ISBI 2017)* 2017 IEEE 14th International Symposium on Biomedical Imaging (ISBI 2017) pp 15–8

Liebgott H, Prost R and Friboulet D 2013 Pre-beamformed RF signal reconstruction in medical ultrasound using compressive sensing *Ultrasonics* **53** 525–33

Lin F, Shelton S E, Espíndola D, Rojas J D, Pinton G and Dayton P A 2017 3-D Ultrasound Localization Microscopy for Identifying Microvascular Morphology Features of Tumor Angiogenesis at a Resolution Beyond the Diffraction Limit of Conventional Ultrasound *Theranostics* **7** 196–204

Liu J, He Q and Luo J 2017 A Compressed Sensing Strategy for Synthetic Transmit Aperture Ultrasound Imaging *IEEE Transactions on Medical Imaging* **36** 878–91

Lorintiu O, Liebgott H, Alessandrini M, Bernard O and Friboulet D 2015 Compressed Sensing Reconstruction of 3D Ultrasound Data Using Dictionary Learning and Line-Wise Subsampling *IEEE Transactions on Medical Imaging* **34** 2467–77

Lustig M, Donoho D and Pauly J M 2007 Sparse MRI: The application of compressed sensing for rapid MR imaging *Magnetic Resonance in Medicine* **58** 1182–95

Montaldo G, Tanter M, Bercoff J, Benech N and Fink M 2009 Coherent plane-wave compounding for very high frame rate ultrasonography and transient elastography *IEEE Transactions on Ultrasonics, Ferroelectrics, and Frequency Control* **56** 489–506

O´Reilly M A and Hynynen K 2013 A super-resolution ultrasound method for brain vascular mapping *Medical Physics* **40** 110701

Opacic T, Dencks S, Theek B, Piepenbrock M, Ackermann D, Rix A, Lammers T, Stickeler E, Delorme S, Schmitz G and Kiessling F 2018 Motion model ultrasound localization microscopy for preclinical and clinical multiparametric tumor characterization *Nature Communications* **9** 1527

Osmanski B, Pernot M, Montaldo G, Bel A, Messas E and Tanter M 2012 Ultrafast Doppler Imaging of Blood Flow Dynamics in the Myocardium *IEEE Transactions on Medical Imaging* **31** 1661–8

Provost J and Lesage F 2009 The Application of Compressed Sensing for Photo-Acoustic Tomography *IEEE Transactions on Medical Imaging* **28** 585–94

Quinsac C, Basarab A and Kouamé D 2012 Frequency Domain Compressive Sampling for Ultrasound Imaging *Advances in Acoustics and Vibration* **2012** 1–16

Ramkumar A and Thittai A K 2020 Strategic Undersampling and Recovery Using Compressed Sensing for Enhancing Ultrasound Image Quality *IEEE Trans. Ultrason., Ferroelect., Freq. Contr.* **67** 547–56

Roux E, Varray F, Petrusca L, Cachard C, Tortoli P and Liebgott H 2018 Experimental 3-D Ultrasound Imaging with 2-D Sparse Arrays using Focused and Diverging Waves *Scientific Reports* **8** 9108

Schiffner M F, Jansen T and Schmitz G 2012 Compressed Sensing for Fast Image Acquisition in Pulse-Echo Ultrasound *Biomedical Engineering / Biomedizinische Technik* **57** Online: https://www.degruyter.com/view/journals/bmte/57/SI-1-Track-B/article-000010151520124142.xml

Sciallero C and Trucco A 2015 Design of a sparse planar array for optimized 3D medical ultrasound imaging *2015 23rd European Signal Processing Conference (EUSIPCO)* 2015 23rd European Signal Processing Conference (EUSIPCO) pp 1341–5

Shahriari S and Garcia D 2018 Meshfree simulations of ultrasound vector flow imaging using smoothed particle hydrodynamics *Phys. Med. Biol.* **63** 205011

Siepmann M, Schmitz G, Bzyl J, Palmowski M and Kiessling F 2011 Imaging tumor vascularity by tracing single microbubbles *2011 IEEE International Ultrasonics Symposium* 2011 IEEE International Ultrasonics Symposium pp 1906–9

Viessmann O M, Eckersley R J, Christensen-Jeffries K, Tang M X and Dunsby C 2013 Acoustic super-resolution with ultrasound and microbubbles *Phys. Med. Biol.* **58** 6447–58

Vilov S, Arnal B, Hojman E, Eldar Y C, Katz O and Bossy E 2020 Super-resolution photoacoustic and ultrasound imaging with sparse arrays *Sci Rep* **10** 4637

Wagner N, Eldar Y C, Feuer A and Friedman Z 2012 Compressed beamforming applied to B-mode ultrasound imaging *2012 9th IEEE International Symposium on Biomedical Imaging (ISBI)* 2012 9th IEEE International Symposium on Biomedical Imaging (ISBI) pp 1080–3

Wang J, Sheng W-X, Han Y-B and Ma X-F 2014 Adaptive beamforming with compressed sensing for sparse receiving array *IEEE Transactions on Aerospace and Electronic Systems* **50** 823–33

Yang G, Yu S, Dong H, Slabaugh G, Dragotti P L, Ye X, Liu F, Arridge S, Keegan J, Guo Y and Firmin D 2018 DAGAN: Deep De-Aliasing Generative Adversarial Networks for Fast Compressed Sensing MRI Reconstruction *IEEE Transactions on Medical Imaging* **37** 1310–21

Yen J T, Steinberg J P and Smith S W 2000 Sparse 2-D array design for real time rectilinear volumetric imaging *IEEE Transactions on Ultrasonics, Ferroelectrics, and Frequency Control* **47** 93–110

Zhang G, Harput S, Hu H, Christensen-Jeffries K, Zhu J, Brown J, Leow C H, Eckersley R J, Dunsby C and Tang M-X 2019 Fast Acoustic Wave Sparsely Activated Localization Microscopy: Ultrasound Super-Resolution Using Plane-Wave Activation of Nanodroplets *IEEE Transactions on Ultrasonics, Ferroelectrics, and Frequency Control* **66** 1039–46

Zhang G, Harput S, Lin S, Christensen-Jeffries K, Leow C H, Brown J, Dunsby C, Eckersley R J and Tang M-X 2018 Acoustic wave sparsely activated localization microscopy (AWSALM):        Super-resolution ultrasound imaging using acoustic activation and deactivation of        nanodroplets *Appl. Phys. Lett.* **113** 014101

Zhang Q, Li B and Shen M 2013 A measurement-domain adaptive beamforming approach for ultrasound instrument based on distributed compressed sensing: Initial development *Ultrasonics* **53** 255–64